\documentclass[10pt,a4paper,aps,prl,superscriptaddress,nofootinbib,nobibnotes,longbibliography,preprintnumbers,floatfix,twocolumn]{revtex4-2}
\usepackage[T1]{fontenc}
\usepackage[utf8]{inputenc}
\usepackage[english]{babel}
\usepackage[dvipsnames, usenames]{xcolor}
\usepackage[unicode, linktocpage, breaklinks,
  colorlinks, citecolor=gray, urlcolor=NavyBlue,
  linkcolor=NavyBlue,
  pdfusetitle]{hyperref}

\usepackage[all]{hypcap}
\usepackage[normalem]{ulem}
\usepackage{csquotes,mathtools}
\usepackage[per-mode = symbol]{siunitx}
\usepackage{physics, tensor, bbm, mathrsfs}
\usepackage{amsmath, amssymb, amsfonts, amsthm}
\usepackage{graphicx, comment, wasysym, placeins}
\usepackage{tikz}
\usetikzlibrary{calc, arrows, arrows.meta, decorations.pathreplacing, 
decorations.pathmorphing,
backgrounds}
\usepackage{orcidlink}
\usepackage{microtype}
\usepackage[nolist,nohyperlinks]{acronym}

\graphicspath{{Figs/}}

\def\RWZ{{\textsc{RWZHyp}\,}}
\def\SpEC{{\textsc{SpEC}\,}}

\allowdisplaybreaks

\begin{document}

\pagenumbering{arabic}

\title{Late-time tails in nonlinear evolutions of merging black holes}

\newcommand{\AEI}{\affiliation{Max Planck Institute for Gravitational Physics (Albert Einstein Institute), Am M\"uhlenberg 1, D-14476 Potsdam, Germany}}
\newcommand{\CornellPhysics}{\affiliation{Department of Physics, Cornell University, Ithaca, NY, 14853, USA}}
\newcommand{\Cornell}{\affiliation{Cornell Center for Astrophysics and Planetary Science, Cornell University, Ithaca, New York 14853, USA}}
\newcommand{\CornellLepp}{\affiliation{Laboratory for Elementary Particle Physics, Cornell University, Ithaca, New York 14853, USA}}
\newcommand{\Caltech}{\affiliation{Theoretical Astrophysics 350-17, California Institute of Technology, Pasadena, CA 91125, USA}}

\author{Marina De Amicis\,\orcidlink{0000-0003-0808-3026}}
\email{marina.de.amicis@nbi.ku.dk}
\affiliation{Niels Bohr International Academy, Niels Bohr Institute, Blegdamsvej 17, 2100 Copenhagen, Denmark}
\author{Hannes Rüter\,\orcidlink{0000-0002-3442-5360}}
\affiliation{CENTRA, Departamento de F\'{\i}sica, Instituto Superior T\'ecnico -- IST, Universidade de Lisboa -- UL,
Avenida Rovisco Pais 1, 1049 Lisboa, Portugal}
\author{Gregorio Carullo\,\orcidlink{0000-0001-9090-1862}}
\email{gregorio.carullo@nbi.ku.dk}
\affiliation{Niels Bohr International Academy, Niels Bohr Institute, Blegdamsvej 17, 2100 Copenhagen, Denmark}
\affiliation{School of Physics and Astronomy and Institute for Gravitational Wave Astronomy, University of Birmingham, Edgbaston, Birmingham, B15 2TT, United Kingdom}
\author{Simone Albanesi\,\orcidlink{0000-0001-7345-4415}}
\affiliation{Theoretisch-Physikalisches Institut, Friedrich-Schiller-Universit{\"a}t Jena, 07743, Jena, Germany}
\affiliation{INFN sezione di Torino, Torino, 10125, Italy}
\author{C. Melize Ferrus\,\orcidlink{0000-0002-2842-2067}}
\affiliation{Theoretical Particle Physics and Cosmology Group, Physics Department, King's  College  London,  Strand,  London  WC2R  2LS,  United Kingdom}
\author{Keefe Mitman\,\orcidlink{0000-0003-0276-3856}}
\Cornell
\author{Leo C. Stein\,\orcidlink{0000-0001-7559-9597}}
\affiliation{Department of Physics and Astronomy,
		University of Mississippi, University, Mississippi 38677, USA}
\author{Vitor Cardoso\,\orcidlink{0000-0003-0553-0433}}
\affiliation{Niels Bohr International Academy, Niels Bohr Institute, Blegdamsvej 17, 2100 Copenhagen, Denmark}
\affiliation{CENTRA, Departamento de F\'{\i}sica, Instituto Superior T\'ecnico -- IST, Universidade de Lisboa -- UL,
Avenida Rovisco Pais 1, 1049 Lisboa, Portugal}

\author{\\ Sebastiano Bernuzzi \orcidlink{0000-0002-2334-0935}}
\affiliation{Theoretisch-Physikalisches Institut, Friedrich-Schiller-Universit{\"a}t Jena, 07743, Jena, Germany}
\author{Michael Boyle \orcidlink{0000-0002-5075-5116}} \Cornell
\author{Nils Deppe \orcidlink{0000-0003-4557-4115}} \CornellLepp \CornellPhysics \Cornell
\author{Lawrence E.~Kidder \orcidlink{0000-0001-5392-7342}} \Cornell
\author{Jordan Moxon \orcidlink{0000-0001-9891-8677}} \Caltech
\author{Alessandro Nagar \orcidlink{}}
\affiliation{INFN sezione di Torino, Torino, 10125, Italy}
\affiliation{Institut des Hautes Etudes Scientifiques, 35 Route de Chartres, Bures-sur-Yvette, 91440, France}
\author{Kyle C.~Nelli \orcidlink{0000-0003-2426-8768}} \Caltech
\author{Harald P. Pfeiffer \orcidlink{0000-0001-9288-519X}} \AEI
\author{Mark A. Scheel \orcidlink{0000-0001-6656-9134}} \Caltech
\author{William Throwe \orcidlink{0000-0001-5059-4378}} \Cornell
\author{Nils L.~Vu \orcidlink{0000-0002-5767-3949}} \Caltech
\author{An\i l Zengino\u{g}lu \orcidlink{0000-0001-7896-6268}}
\affiliation{Institute for Physical Science \& Technology, University of Maryland, \\ College Park, MD 20742-2431, USA}

\hypersetup{pdfauthor={De Amicis et al.}}

\begin{abstract}
We uncover late-time gravitational-wave tails in fully nonlinear 3+1 dimensional numerical relativity simulations of merging black holes, using the highly accurate \SpEC code.
We achieve this result by exploiting the strong magnification of late-time tails due to binary eccentricity, recently observed in perturbative evolutions, and showcase here the tail presence in head-on configurations for several mass ratios close to unity.
We validate the result through a large battery of numerical tests and detailed comparison with perturbative evolutions, which display striking agreement with full nonlinear ones.
Our results offer yet another confirmation of the highly predictive power of black hole perturbation theory in the presence of a source, even when applied to nonlinear solutions.
The late-time tail signal is much more prominent than anticipated until recently, and possibly within reach of gravitational-wave detectors measurements, unlocking observational investigations of an additional set of general relativistic predictions on the long-range gravitational dynamics.
\end{abstract}

\maketitle

\noindent {\textbf{\textit{Introduction}}.}
%
%
In the last few decades, the merger dynamics and subsequent relaxation regime (quasinormal ``ringing'') of compact binaries with comparable-masses have been the subject of considerable modeling efforts, chiefly enabled by fully nonlinear numerical relativity (NR) evolutions~\cite{Pretorius:2005gq,Campanelli:2005dd,Baker:2005vv,Berti:2007fi}.
Gravitational-wave (GW) observations of coalescing black hole (BH) binaries~\cite{LIGOScientific:2016aoc,AdvLIGO,AdvVirgo,Kagra}, granting an unprecedented opportunity of accessing the highly nonlinear regime of dynamical spacetimes, were the main driver
for these efforts.
In contrast, the subsequent late-time dynamics received much less attention from the physics community, partly because its signatures were thought to be out of reach of foreseeable experimental efforts.

%
%
BH perturbation theory~\cite{1960AnPhy...9..220D} predicts that
the dominant late-time effect in the relaxation of compact objects 
is a power-law contribution (``tail''),
both in spherical symmetry~\cite{Price:1971fb,Price:1972pw,Cunningham:1978zfa,gomez1992asymptotics,Gundlach:1993tp,Gundlach:1993tn,Leaver:1986gd,Ching:1994bd,Andersson:1996cm,Burko:1997tb,Barack:1998bw,Bernuzzi:2008rq,Hod:2009my} and 
axisymmetry~\cite{Krivan:1996da,Krivan:1997hc,Hod:1999ci,Hod:1999rx,Barack:1999ma,Ori:1999nc,Krivan:1999wh,Poisson:2002jz,Burko:2002bt,Burko:2004jn,Burko:2007ju,Burko:2010zj,Racz:2011qu,Zenginoglu:2012us,Burko:2013bra,Harms:2013ib}.
For instance, at future null infinity ($\mathcal{I}^+$), 
metric perturbations of a Schwarzschild BH decay with
the leading-order power-law $u^{-(\ell+2)}$, with $u$ the retarded time and $\ell$ the waveform multipole.\footnote{An especially clear discussion of the spherically symmetric case can be found in Refs.~\cite{Leaver:1986gd,Andersson:1996cm}, and Chapter 12 of Ref.~\cite{Maggiore:2018sht}.}
%
%
%
Tails constitute a clean probe of the
long-range structure of dynamical spacetimes
and allow for distinct predictions of the 
relaxation of compact object within GR~\cite{Hod:1999ci}.
Even more interestingly, as shown in~\cite{DeAmicis:2024not} and discussed below, source-driven tails at intermediate times bear direct imprints of the \textit{strong-field binary dynamics}.
So far, late-time tails have only been studied numerically in a perturbative framework~\cite{Cunningham:1978zfa,Price:1994pm,Krivan:1997hc,Bernuzzi:2008rq,Harms:2013ib, Albanesi:2023bgi, DeAmicis:2024not, Islam:2024vro}, but never observed in nonlinear NR simulations of comparable-mass BH mergers.
This was mostly due to the very high accuracy required to unveil them, together with the need for solid control over boundary and initial data (ID) effects.
Here we present the first robust extraction of 
late-time tail terms from fully nonlinear NR simulations of comparable-mass BH binaries.

\begin{figure}[t!]
\centering
\includegraphics[width=0.49\textwidth]{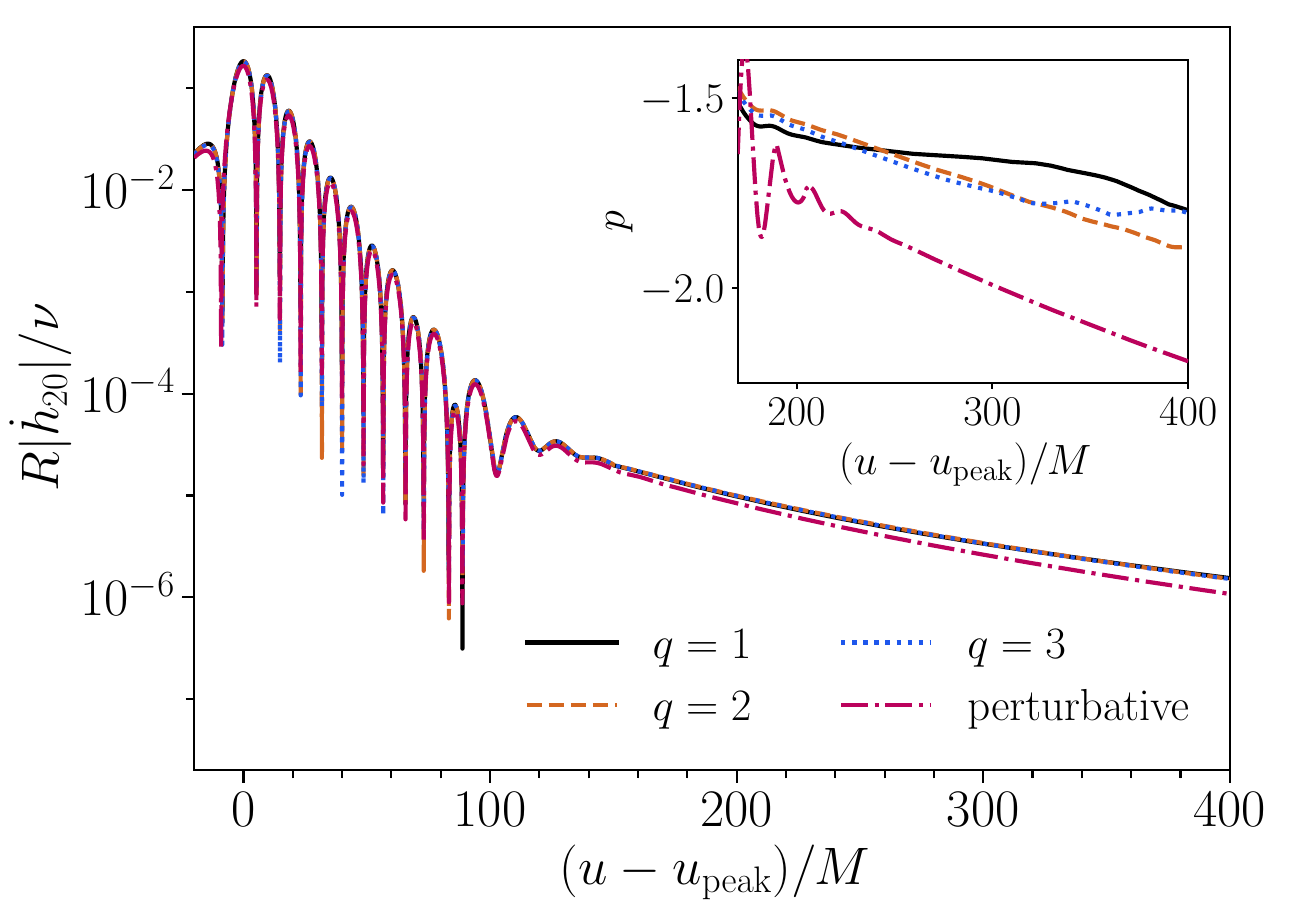}
\caption{
Mass-rescaled quadrupolar news amplitude, as a function of retarded time from the peak.
The inset shows the behaviour of the tail exponent, as defined in Eq.~\eqref{eq:tail_exponent}.
Thick lines represent nonlinear evolution of head-on comparable-mass collisions, while the dot-dashed line represents a perturbative evolution of a radial infall with compatible initial data.
}
\label{fig:SXS_RWZ}
\end{figure}

Characterising tail effects in numerical simulations is challenging, because excitations of quasinormal-modes (QNM) can take a long time 
to decay~\cite{Harms:2013ib}.
Hence, to identify tail contributions we must look for them in a regime where QNMs are short-lived, and the tail amplitude is large.
The first is easy to achieve by targeting remnant BHs with small angular momentum (shorter relaxation time).
The second condition is harder to obtain, because we do not know a-priori the dependence of the tail amplitude on the binary's initial conditions.
Recently, some of us~\cite{Albanesi:2023bgi} found through 
effective-one-body (EOB)~\cite{Buonanno:1998gg, Buonanno:2000ef,Damour:2007xr} calculations in the small-mass-ratio limit, that
the binary eccentricity induces a strong enhancement of the tail amplitude.
Tails in perturbative radial infalls, with a similar source-driven setup that we employ, were computed in e.\,g.~\cite{Nagar:2006xv} and for the first time at $\mathcal{I}^+$ in~\cite{Bernuzzi:2012ku}.
The mechanism behind this enhancement, and whether its explanation necessitates a modification to Price's law were initially unknown.
In~\cite{DeAmicis:2024not} some of us subsequently constructed an analytical model, that accurately describes the source-driven tail behavior in test-particle evolutions (for a given trajectory).
The model shows how the simple picture of a single ``Price tail'' needs to be replaced by a superposition of a large number of inverse power-law components giving rise to a long slowly-decaying transient (see also~\cite{Islam:2024vro}).
A key prediction of the model is that the tail amplitude reaches its maximum for radial infalls, hence in this work we
restrict to head-on collisions.
These analytical predictions, together with the considerations presented in~\cite{Allen:2004js,Dafermos:2004wt}, equip us with the necessary understanding of ID and boundary effects to attempt extracting a tail signal in nonlinear evolution, and constitute the required semi-analytical tool to verify the robustness of numerical evolutions.

Exploiting the very high accuracy of the \SpEC code~\cite{SpECwebsite}, we demonstrate the extraction of tail effects in fully nonlinear 3+1  evolutions, displayed in Fig.~\ref{fig:SXS_RWZ}.
The figure shows the amplitude of the GW news quadrupolar mode extrapolated to $\mathcal{I}^+$, for several head-on binary simulations with mass ratios close to unity.
Around $140M$ after the peak, the amplitude transitions from an 
exponentially damped
quasinormal-driven regime to a slowly-decaying non-oscillatory
behaviour.
The result aligns remarkably well with perturbative linear evolutions of an infalling test particle with compatible ID,
further validating the numerical computation and pointing towards a suppression of nonlinear corrections~\cite{Cardoso:2024jme}.

Below, we report on the numerical methods employed both in the nonlinear and perturbative cases.
We discuss in more detail the results obtained, their interpretation, and possible nonlinear effects behind the subtle differences between linear and nonlinear evolutions.
We conclude summarising open questions and future detectability avenues for tail signals.
In the End Matter we give further details of the linearised analytical computation, and present a series of tests to confirm the robustness of our numerical results.\\

\noindent {\textbf{\textit{Conventions}}.}
%
We use geometric units $c=G=1$.
The GW strain is decomposed in spin-weight $-2$ spherical harmonics modes, $h_{\ell m}(t)$.
To avoid memory contributions or gauge effects that could spoil tail extraction, entering as a constant offset at late times, we focus on the GW news function $\dot{h}_{\ell m}$.
Modes beyond the quadrupolar are subdominant, hence we are going to focus on $\ell=2$.
Exploiting the cylindrical symmetry of the problem, we present all results in a frame~\cite{Gualtieri:2008ux} in which the two BHs collide along the $z$-axis and hence $(\ell,m) = (2,0)$ is the only non-zero $\ell=2$ waveform multipole.
At asymptotically late-times, it holds:  $h_{\ell m} \propto u^{\bar{p}}, \,\dot{h}_{\ell m} \propto u^{\bar{p}-1}$,
with $\bar{p} = -(\ell+2)$ for Schwarzschild BHs.
Hence it is convenient to define a ``tail exponent'' as
\begin{equation}
    p(t) = 1+\frac{d\ln|\dot{h}_{\ell m}|}{d\ln u} \, .
    \label{eq:tail_exponent}
\end{equation}
so that $p(t) = \bar{p}$ at late times.
With $m_{1,2}$ we indicate the Christodoulou masses~\cite{Christodoulou71} 
of the individual black holes at the relaxation time of the
simulation (i.e., when the high-frequency oscillations in the Christodoulou masses have settled down), $M=m_1 + m_2$ is the total Christodoulou mass of the system, $\nu=m_1m_2/M^2$ the symmetric mass ratio, and $q=m_1/m_2 \ge 1$ the binary mass ratio.
The time axis is constructed by setting $t=0$ at the peak of $|\dot{h}_{20}(t)|$, and is quoted in units of $M$.
The tortoise coordinate in perturbative evolutions is $r_*=r+2M\log(\frac{r}{2M}-1)$, with $r$ the standard Schwarzschild coordinate.
We indicate the distance between the binary and the observer with $R$.\\

\noindent {\textbf{\textit{Numerical methods}}.}
%
We generate non-linear evolutions of head-on collisions with the 
\SpEC code~\cite{Scheel:2008rj,Szilagyi:2009qz,Hemberger:2012jz,Ossokine:2013zga,SpECwebsite}, 
whose methods are summarised in~\cite{Mrou__2013,Boyle:2019kee}.
The ID~\cite{Lovelace:2008tw,Buchman:2012dw,Cook:2004kt,Pfeiffer:2002wt} are constructed using the extended
conformal thin sandwich equations~\cite{Pfeiffer2003}
and the
evolution is carried out with the generalized harmonic
formulation~\cite{Lindblom:2005qh,Lindblom_2009}.
Standard {\SpEC}~runs stop at a retarded time of $100 M$ after merger,
which is typically enough to only resolve the quasinormal ringdown.
In order to capture the late-time waveform behavior, here we carry
out simulations with much longer post-merger components, stopping at retarded time $500 M$ after merger.

We compare these NR waveforms with linear perturbative waveforms of radial infalls towards a Schwarzschild BH, numerically computed using the \RWZ code~\cite{Bernuzzi:2010ty,Bernuzzi:2011aj}.
We solve for the Regge-Wheeler/Zerilli equations, governing the evolution in time of gauge invariant quantities $\Psi^{(e/o)}_{\ell m}$
\begin{equation}
        \left[\partial_t^2 -\partial_{r_*}^2+V_{\ell m}^{(e/o)}(r_*)\right]\Psi_{\ell m}^{(e/o)}(t,r_*) = S_{\ell m}^{(e/o)}(t,r) \, .
    \label{eq:RWZ_equation}
\end{equation}
These master variables encode the odd/even metric degrees of freedom at the linear level~\cite{Nagar:2006xv}.
In the linear evolutions, we set null ID $\Psi^{(e/o)}_{\ell m}(t=0,r)=\partial_t \Psi^{(e/o)}_{\ell m}(t=0,r)=0$, as the perturbations are driven by the presence of the source $S^{(e/o)}_{\ell m}$.
The latter is evaluated on the infalling test-particle's trajectory.
We refer to~\cite{DeAmicis:2024not} for additional details and the source explicit expression.
For radial infalls with null ID, the odd sector is identically zero, hence in what follows we drop the superscript.
Following an EOB-inspired approach, we compute the particle's trajectory solving the associated Hamiltonian equations of motion (see~\cite{Albanesi:2023bgi} for the explicit expressions), where dissipative effects linked to the GW emission are taken into account by including a radiation reaction force~\cite{Nagar:2006xv,Damour:2007xr}. 
Such a force was computed, for generic orbits by means of a PN-based, EOB-resummed analytical expansion~\cite{Bini:2012ji,Chiaramello:2020ehz} for the fluxes of energy and angular momentum observed at infinity, and were shown to be consistent with the corresponding numerical quantities in Ref.~\cite{Albanesi:2021rby}.
%
%
Note that, due to the short time-scale of our dynamics, the inclusion of the radiation reaction force has a small impact on the evolution.

At large distances, the master function $\Psi_{\ell m}$ is related to the linearized strain's spin weight $-2$ spherical harmonic modes $h_{\ell m}$ through~\cite{Nagar:2005ea}
\begin{equation}
h_{\ell m}=\dfrac{1}{r} \sqrt{\dfrac{\left(\ell+2\right)!}{\left(\ell-2\right)!}} \, \Psi_{\ell  m}+\mathcal{O}\left(\dfrac{1}{r^2}\right) \, .
\end{equation}
Moreover, the source term $S_{\ell  m}(t,r)$ is linearly proportional to the mass ratio, that in the test-particle limit is equivalent to the symmetric mass ratio, so $\Psi_{\ell m}\propto \nu$. 
To compare perturbative waveforms with NR comparable-mass results, we will thus always rescale the perturbative results $\Psi_{\ell m}$ by $\nu$.
The NR waveforms are also rescaled with their symmetric mass-ratio.

%
To compare full NR results against perturbative test-mass limit ones, we initialise the two systems with compatible ID.
We only consider non-spinning black holes.
The SXS simulation ID are given by setting the angular momentum to
zero and by imposing that the Arnowitt-Deser-Misner (ADM) energy
$E_\mathrm{ADM}$ is equal to the total rest mass of the system $M$
within relative accuracy of $10^{-4}$, so that the two black holes are at rest at
infinite separation, and the initial binding energy is close to zero.
We generate equivalent test-mass data by imposing that the test-particle is at rest at infinity, i.e.\ by setting its initial energy equal to its mass.\\
%

\noindent {\textbf{\textit{Outer Boundary}}.}
%
\begin{figure}[tb]
  \begin{center}
    \begin{tikzpicture}[scale=0.45,>=Stealth]

      \newcommand{\Rext}{4}
      \newcommand{\Rout}{15}
      \newcommand{\Tmax}{12}
      \newcommand{\Tmerg}{3}
      \pgfmathsetmacro{\Rnought}{(\Tmerg)^(2/3)}

      \draw[->] (-0.5,0) -- (1+\Rout,0) node[right] {$r$};
      \draw[->] (0,-0.5) -- (0,\Tmax) node[above] {$t$};
      \draw[color=blue, very thick, variable=\t, domain=0:\Tmerg, samples=100]
      plot ({(\Tmerg-\t)^(2/3)},{\t});

      \draw[color=blue, very thick] (0, \Tmerg) -- (0, \Tmax);

      \draw[color=red, dashed] (0,0) --
      node[pos=0.6, above, sloped] {$u=0$}
      (\Rext,\Rext);

      \draw[black] (\Rext,0) node[below] {$R_{\text{ext}}$} -- (\Rext,\Tmax);
      \draw[black] (\Rout,0) node[below] {$R_{\text{out}}$} -- (\Rout,\Tmax);

      \draw[black] (0,\Tmerg) node[left] {$T_{\text{merg}}$};

      \draw[color=red, dashed] (\Rout, 0) --
      node[midway, above, sloped] {causal contact with boundary}
      (\Rext, \Rout - \Rext);

      \draw[black, <->] (-0.5+\Rext,\Rext) --
      node[midway, above, sloped] {$R_{\text{out}}-2R_{\text{ext}}$}
      (-0.5+\Rext, \Rout - \Rext);

      \draw (\Rnought,0) node[below] {$R_{0}$};

      \draw[color=ForestGreen, ->, decorate, 
      decoration={snake, 
      segment length=15,
      post length=1pt}]
      ($(0,\Tmerg)!0.1!(\Rext, \Tmerg+\Rext)$) 
      -- ($(0,\Tmerg)!0.65!(\Rext, \Tmerg+\Rext)$);

    \end{tikzpicture}
  \end{center}

  \caption{%
    Schematic spacetime diagram demonstrating the geometry of
    the head-on merger and remnant, retarded time $u$, extraction spheres, and
    outer boundary.
  }
  \label{fig:boundary-obs_causal_contact_visualization}
\end{figure}
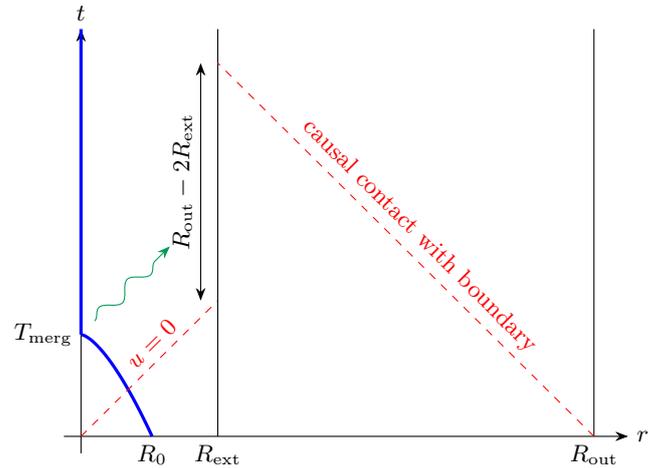
The \SpEC code imposes data on the outer boundary, located on a sphere with 
radius~$R_\mathrm{out}$, such that
constraints are preserved~\cite{Lindblom:2005qh}, and the physical
degrees of freedom are chosen assuming that there is no 
gravitational radiation entering the numerical
domain~\cite{Rinne_2006,Rinne_2007}. 
In curved spacetimes, however, a certain amount of radiation is always scattered back.
Back-scattering, giving rise to wave propagation well within the light-cone, is precisely the mechanism responsible for tails generation~\cite{Blanchet:1992br, Blanchet:1993ec,Poisson:1993vp, Poisson:1994yf,DeAmicis:2024not}, and a small $R_\mathrm{out}$ would imply missing a significant fraction of such contribution.
Hence, to capture the vast majority of the back-scattered
radiation, we place the outer boundary exceptionally far away from the binary location (see Tab.~\ref{tab:sims_ecc}).

Moreover, due to imperfect boundary conditions, when radiation reaches the boundary, a non-physical numerical artifact is generated, contaminating the signal with numerical noise.
As pointed out in Refs.~\cite{Allen:2004js,Dafermos:2004wt}, this contamination can alter the structure of tails. Therefore, to study the long-range and late-time tails contribution, it is important that our simulations remain causally disconnected from the boundary.
To avoid such contamination, we chose $R_{\rm out}$ to be large enough such that the boundary is never in causal contact with the extraction spheres for the entire evolution (see Table~\ref{tab:sims_ecc}).
This causal structure, our evolution domain, and the locations of finite-radius observers are visualized in Fig.~\ref{fig:boundary-obs_causal_contact_visualization}.
In the End Matter, in Table~\ref{tab:sims_suppl} we
report the details of additional \SpEC simulations in which there
is causal contact between the boundary and the extraction spheres,
clearly showing a strong impact on the behavior of the tail.

\noindent {\textbf{\textit{GW extraction}}.}
%
The \RWZ software uses a numerical domain that is decomposed
into two regions, a compact inner region and 
an outer hyperboloidal layer~\cite{Zenginoglu:2007jw,Zenginoglu:2009ey,Zenginoglu:2010cq}.
The inner region contains the particle trajectories and
has a uniform grid in $r_*$, ranging from 
$r_* = -100M$ to large a value of $r_*$.
The hyperboloidal layer, compactified over $r_*$,
is attached at the end of the inner
region in order to bring $\mathcal{I}^+$ into the computational domain.
As a result, perturbative waveforms can be computed both at finite distances, in terms of the coordinate time $t$, as well as at $\mathcal{I}^+$ in terms of the retarded time $u$.

In \SpEC the GW information is extracted on spheres, whose radii $R_{\text{ext}}$ are 
distributed in the interval $[300M,1200M]$ and are held
fixed across all simulations in this work, as opposed to
standard {\SpEC}~runs in which their radii scale with $R_\mathrm{out}$.
The extracted finite radius waveform data are extrapolated to $\mathcal{I}^+$
by a standard polynomial fit using the method 
of~\cite{Iozzo21} as implemented in the
\textsc{scri} package~\cite{scri,scrirepo,Boyle2013,BoyleEtAl:2014,Boyle2015a}.

We give additional details of the simulations in the End Matter, where we present convergence tests in terms of resolution, extrapolation methods, and initial separation, validating the robustness of our results. Furthermore we discuss results from waveform extraction by Cauchy-characteristic evolution.
\\

\noindent {\textbf{\textit{Results}}.}
%
\begin{table}[t!]
\begin{center}
\begin{tabular}{ccccc}
\hline \hline \noalign{\smallskip}
SXS identifier & $q$ & $D_0/M$ & $R_\mathrm{out}/M$ & $u_{con}/M$ \\
\noalign{\smallskip}\hline \noalign{\smallskip}
SXS:BBH:3991 & 1 & 100 & 4000  & 1600 \\
SXS:BBH:3997 & 2 & 100 & 4000  & 1600 \\
SXS:BBH:3998 & 3 & 100 & 4000  & 1600 \\
SXS:BBH:3994 & 1 & 100 & 8000  & 5600 \\
SXS:BBH:3995 & 1 & 200 & 8000  & 5600 \\
SXS:BBH:3996 & 1 & 400 & 8000  & 5600 \\
%
%
\noalign{\smallskip}\hline \hline
\end{tabular}
\caption{List of simulations and relevant parameters:
$D_0$ is the initial separation,
$R_\mathrm{out}$ the radius of the outer boundary and
$u_{con} = R_\mathrm{out} - 2 R_\mathrm{ext}$ denotes the approximate retarded time at which
the outermost extraction radius (located at $1200M$) would enter in
causal contact with the outer boundary.
The first column states the identifier in the SXS waveform
catalog~\cite{Boyle:2019kee,SXScatalog}.
}
\label{tab:sims_ecc}
\end{center}
\end{table}
%
%
In Table~\ref{tab:sims_ecc} we report the parameters of the simulations carried out with the above methods.
Fig.~\ref{fig:SXS_RWZ} shows the resulting news function, normalized by the symmetric mass ratio, with respect to the retarded time $u$.
%
%
%
A first striking result from Fig.~\ref{fig:SXS_RWZ} concerns the dependence of the tail on the mass ratio.
In fact, after mass-rescaling the waveforms appropriately, the comparable mass waveforms with different mass ratio are very similar (around the percent level), suggesting that finite mass ratio corrections do not play a significant role in the waveform generation, including the tail part.

We also compute the tail exponent $p$ (Eq.~\eqref{eq:tail_exponent}), reported in Fig.~\ref{fig:SXS_RWZ} inset.
Its magnitude is much smaller than the asymptotic Price-law value ($\bar{p} 
= -4$, for this multipole), towards which we would expect it to slowly converge~\cite{DeAmicis:2024not}.
%
Such decrease in exponent magnitude significantly boosts the tail amplitude at intermediate times.
This result agrees with the analytical perturbative picture of Ref. ~\cite{DeAmicis:2024not}, according to which tail emission is maximized for motion at large distances from the central BH, with small angular velocity.
The additional numerical time derivative required for the exponent computation, combined with finite resolution, tends to introduce high-frequency noise.
To compute the tail exponent at late-times we therefore apply a Savitzky-Golay filter~\cite{Savitzky1964} on the waveform to suppress high frequency oscillations.
In the End Matter we compare the unfiltered tail exponent with respect to the one computed after applying the Savitzky-Golay filter, showcasing that our conclusions are not impacted by the filtering.

%
Even the test-mass perturbative case (similarly rescaled, and aligned minimising the post-peak mismatch) shows a remarkable agreement with the nonlinear evolutions, displaying the same slowly-decaying behaviour and an identical overall morphology.
Such results confirm that the tail is primarily generated by the source term.
This aligns with previous findings, which showcased how test-particle perturbative evolutions proved to be a remarkably accurate tool for the modelling of inspiral-merger-ringdown waveforms~\cite{Nagar:2006xv,Damour:2007xr,Barausse:2011kb}.
Somewhat surprisingly though, this framework provides \textit{quantitatively accurate} predictions even for comparable mass systems~\cite{Wardell:2021fyy, Islam:2023aec}, as we confirm here.
This is true even in the (a priori strong-field) merger stage, as notably depicted also in Fig.~2 of~\cite{Nagar:2022icd}.

%
%
However, in Fig.~\ref{fig:SXS_RWZ} above there is visibly growing mismatch as the tail evolves, with the nonlinear evolution displaying a slower decay with respect to the perturbative result.
This may be due to a variety of effects, including the presence of nonlinear tail components (computed in~\cite{Cardoso:2024jme} for $\ell=4$ modes within second order perturbation theory, see also~\cite{Okuzumi:2008ej}), together with corrections due the finite mass-ratio or the time-dependent background nature in the nonlinear case~\cite{Sberna:2021eui,Redondo-Yuste:2023ipg,Zhu:2024dyl,May:2024rrg,Capuano:2024qhv}.
Although the good agreement in the quasinormal-driven regime suggests the last two effects are likely small, the hereditary nature of tails implies that small differences in the evolution can accumulate and impart a larger effect.\\

\noindent {\textbf{\textit{Future avenues}}.}
%
We have uncovered late-time gravitational-wave tails in 3+1 nonlinear simulations of black holes collisions through the highly accurate \SpEC code,  robustly validating this result through a number of numerical tests.
The waveforms display a remarkable agreement with perturbative results, a fact that deserves further scrutiny in future works.

Our results raise several interesting questions.
On the modeling side, these include:
What is the non-linear content of late-time tails? 
Can second-order tails explain the observed differences between linear and nonlinear evolutions?
Or can these instead be accounted for by higher-order (self-force type) corrections to the trajectory or the dynamical background?
Precision studies on this subject will benefit from longer simulations and the development of Cauchy Characteristic Matching techniques~\cite{Bishop:1998uk,Szilagyi:2000xu,Ma:2023qjn} or non-linear codes in hyperboloidal coordinates~\cite{Peterson:2024bxk}, to uncover the subtle role of nonlinearities.

On the observational side, the tail sensitivity to the long-range spacetime structure implies it might be used as a tool to detect ``environmental'' effects around binaries, such as those induced by dark matter halos, accretion disks, boson clouds, as well as a tertiary object.
This could open up an entire new way of characterising non-vacuum spacetimes, as displayed e.g.\ in~\cite{Spieksma:2024voy} and discussed in~\cite{Cardoso:2024jme}.
The impact of these configurations on tails will be the subject of future investigation.
Observability studies will greatly benefit from the construction of analytical tail models for spinning remnants (akin to~\cite{DeAmicis:2024not}), as spin might enhance tails even further~\cite{Islam:2024vro}.

The source-driven enhancement of the tail brings it much closer to the reach of upcoming GW observatories, hence our results open the possibility of observational tests of general relativistic predictions on the long-range structure of highly-curved spacetimes.
A detailed tail detectability study is the subject of ongoing investigations.\\


\noindent {\textbf{\textit{Acknowledgments}}.}
%
%
M. De A. expresses her gratitude to Perimeter Institute for its kind hospitality during the final stages of the paper's preparation.
M. De A. also thanks Sizheng Ma and Luis Lehner for stimulating discussions on the topic during the visit.
S.A. gratefully acknowledges the warm hospitality of the Niels Bohr Institute, where this work was initiated.
G.C. acknowledges funding from the European Union’s Horizon 2020 research and innovation program under the Marie Sk{\l}odowska-Curie grant agreement No. 847523 ‘INTERACTIONS’.
We acknowledge support from the Villum Investigator program by the VILLUM Foundation (grant no. VIL37766) and the DNRF Chair program (grant no. DNRF162) by the Danish National Research Foundation.
H.R.R. and V.~C. acknowledge financial support provided under the European Union’s H2020 ERC Advanced Grant ``Black holes: gravitational engines of discovery'' grant agreement no. Gravitas–101052587.
Views and opinions expressed are however those of the authors only and do not necessarily reflect those of the European Union or the European Research Council. Neither the European Union nor the granting authority can be held responsible for them.
This project has received funding from the European Union's Horizon 2020 research and innovation programme under the Marie Sk{\l}odowska-Curie grant agreement No 101007855 and No 101131233.\\
K.M. is supported by NASA through the NASA Hubble Fellowship grant \#HST-HF2-51562.001-A awarded by the Space Telescope Science Institute, which is operated by the Association of Universities for Research in Astronomy, Incorporated, under NASA contract NAS5-26555.
S.A. acknowledges support from the Deutsche Forschungsgemeinschaft (DFG) project ``GROOVHY'' 
(BE 6301/5-1 Projektnummer: 523180871).
L.C.S. acknowledges support from NSF CAREER Award PHY--2047382 and a Sloan Foundation Research Fellowship.\\
%
This material is based upon work supported by the National Science Foundation under Grants No.~PHY-2407742, No.~PHY-2207342, and No.~OAC-2209655 at Cornell. Any opinions, findings, and conclusions or recommendations expressed in this material are those of the author(s) and do not necessarily reflect the views of the National Science Foundation. This work was supported by the Sherman Fairchild Foundation at Cornell.
%
This work was supported in part by the Sherman Fairchild Foundation and by NSF Grants No.~PHY-2309211, No.~PHY-2309231, and No.~OAC-2209656 at Caltech.
Simulations have been carried out on the Frontera computing system
within the Texas Advanced Computing Center (TACC) grant PHY20018,
``Gravitational Waves from Compact Binaries: Computational
Contributions to LIGO''.

\noindent{\textbf{\textit{Software}}.}
The version of the \RWZ \, code used bears the tag \texttt{tails}, on the  \texttt{rwzhyp{\_}eccentric} branch.
The manuscript content has been derived using publicly available software: 
\textsc{matplotlib}~\cite{matplotlib}, 
\textsc{numpy}~\cite{numpy}, 
\textsc{scipy}~\cite{scipy} 
and \textsc{sxs}~\cite{Boyle_The_sxs_package_2023}.

\noindent{\textbf{\textit{Data availability}}.}
\SpEC simulation data will be made available in the SXS waveform catalog~\cite{SXScatalog}.

\bibliography{Bibliography}

\clearpage

\section*{END MATTER}

\begin{table}[t!]
\begin{center}
\begin{tabular}{ccccr}
\hline \hline \noalign{\smallskip}
SXS identifier & $q$ & $D_0/M$ & $R_\mathrm{out}/M$ & $u_{con}/M$ \\
\noalign{\smallskip}\hline \noalign{\smallskip}
SXS:BBH:3985 & 1 & 100 & 1300  &  $-1100$ \\
SXS:BBH:3986 & 1 & 200 & 1300  &  $-1100$ \\
SXS:BBH:3987 & 1 & 400 & 1300  &  $-1100$ \\
SXS:BBH:3988 & 1 & 100 & 2300  &  $-100$ \\
SXS:BBH:3989 & 1 & 200 & 2300  &  $-100$ \\
SXS:BBH:3990 & 1 & 400 & 2300  &  $-100$ \\
SXS:BBH:3992 & 1 & 200 & 4000  &  1600 \\
SXS:BBH:3993 & 1 & 400 & 4000  &  1600 \\
%
%
 \noalign{\smallskip} \hline \hline
\end{tabular}
\caption{
Parameters as in table~\ref{tab:sims_ecc} for additional 
simulations with the same setup as other simulations in this work,
but with the causal contact between extraction spheres and boundary.}
\label{tab:sims_suppl}
\end{center}
\end{table}

\noindent {\bf \em Impact of initial separation.} 
%
\begin{figure}[b!]
\includegraphics[width=\columnwidth]{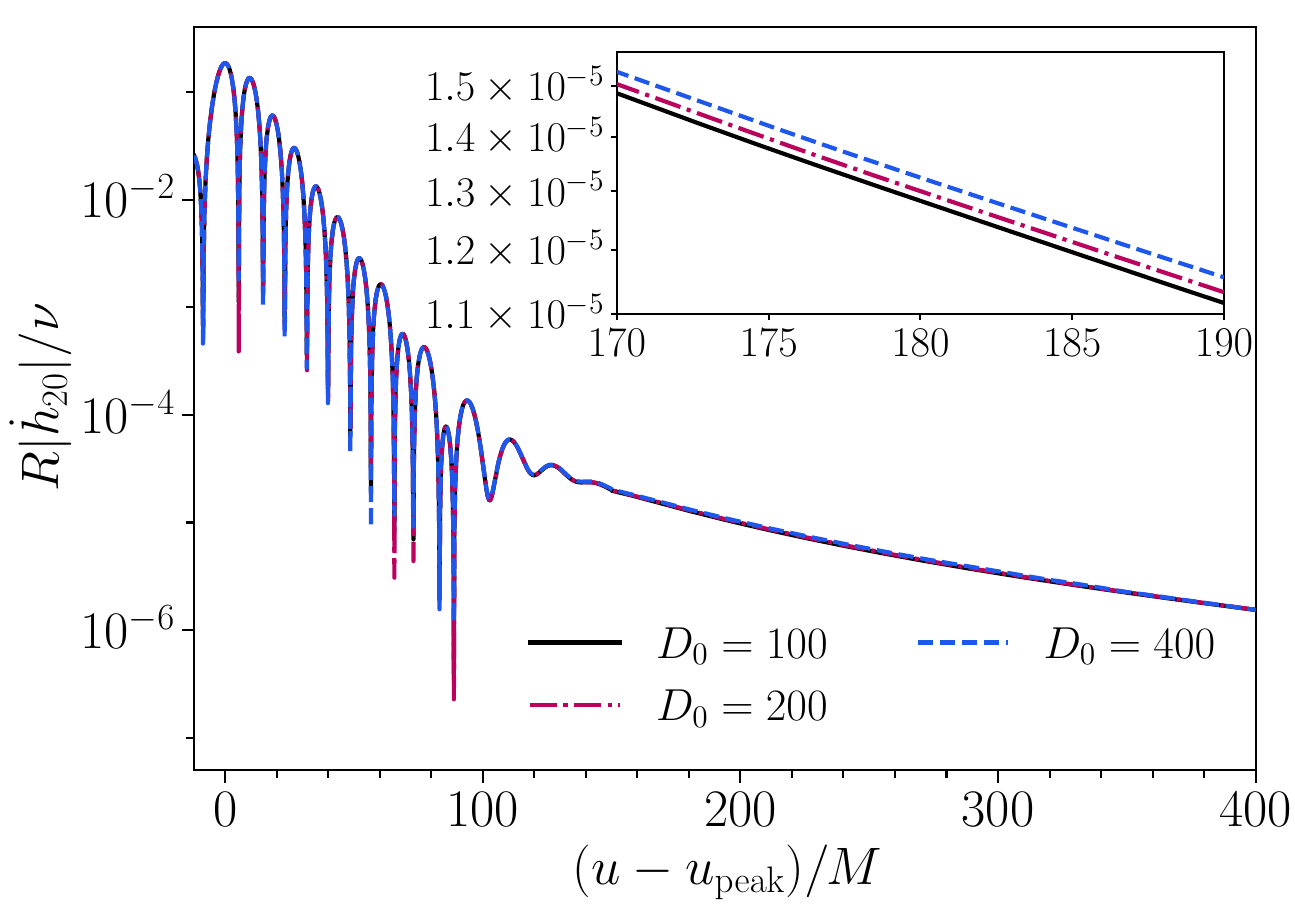}
\caption{Equal-mass case for different initial separations.
\label{fig:SXS_D0}}
\end{figure}
We investigate a sequence of NR simulations with $q=1$ and varying initial separation (with identifiers SXS:BBH:3994, SXS:BBH:3995 and SXS:BBH:3996 in Table~\ref{tab:sims_ecc}).
If ID are set up consistently, the three evolutions should give close to indistinguishable results.
This is what Fig.~\ref{fig:SXS_D0} shows, with very small differences even when increasing the initial separation by a factor of four.
This confirms the robustness of our numerical evolutions.
However, in practice our ID at different separation will not be perfect (i.e.\ will not correctly capture the entire past binary history).
In this case, the tail amplitude is expected to \textit{increase} with larger initial separation, since this enhances the overlap between the source and the tail propagator~\cite{DeAmicis:2024not}, consistently with tails hereditary nature~\cite{Blanchet:1992br, Blanchet:1993ec, Blanchet:1994ez,Poisson:1993vp, Poisson:1994yf}.
Such increase is expected to be small if we are already at large enough separations, so that most of the relevant binary history is already captured.
This is confirmed by Fig.~\ref{fig:SXS_D0}, showing a small increase of the tail amplitude when increasing the separation, another validation of the robustness of our results.\\

\begin{figure}[b!]
\centering
\includegraphics[width=\columnwidth]{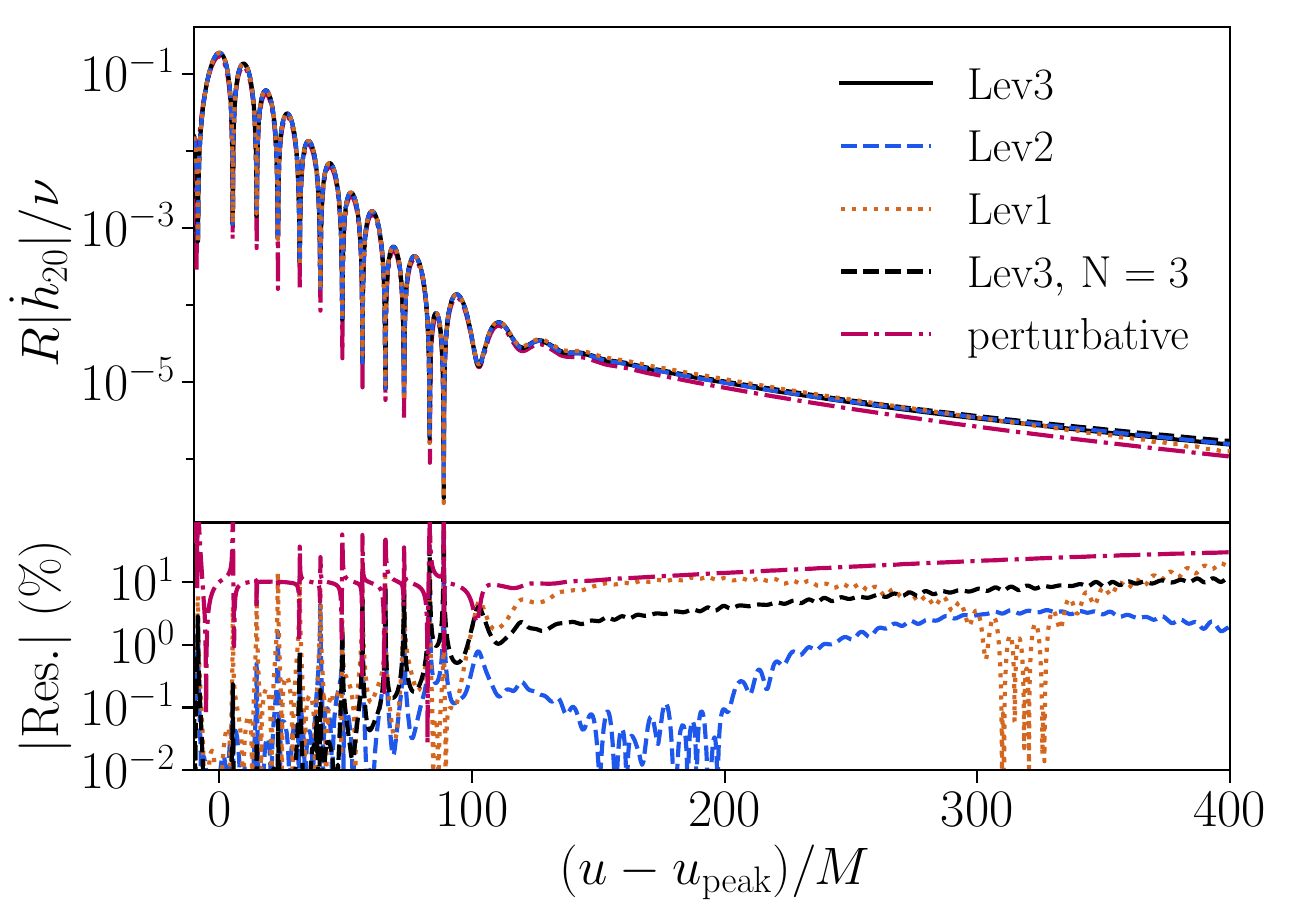}
\caption{Top: different resolutions levels (solid) for run SXS:BBH:3991.
A higher extrapolation order is shown in dashed black (other lines refer to $N=2$), while the purple dot-dashed line reports the perturbative result.
Bottom: residuals with respect to the highest resolution ($\mathrm{Lev 3}$).
}
\label{fig:resolution}
\end{figure}

\noindent {\bf \em Resolution tests.} 
%
In Fig.~\ref{fig:resolution} we compare three different resolution levels (see~\cite{Mrou__2013,Boyle:2019kee}), for the waveform of run SXS:BBH:3991 (see Tab.~\ref{tab:sims_ecc}).
We report the residuals relative to the highest resolution available, $\mathrm{Lev}3$ with smaller extrapolation order, $N=2$, defined as
\begin{equation}
    \mathrm{Res.}({\rm X})=100\frac{|\dot{h}^{(\rm X)}_{20}|-|\dot{h}^{(\rm Lev3,N=2)}_{20}|}{|\dot{h}^{(\rm Lev3,N=2)}_{20}|} \, .
    \label{eq:res_def}
\end{equation}
The tail properties are unchanged with increasing resolution, and differences between different resolution levels are too small to affect any of the considerations reported in the main text, including comparisons with perturbative waveforms.\\

\noindent {\bf \em Extrapolation procedure.} 
%
We first recall known perturbative predictions to gain intuition on waveform extrapolation, and then test our results robustness with respect to the extrapolation procedure employed in the main text.

\textit{Perturbative picture --}
The tail in the strain observed at finite distances in terms of the coordinate time $t$ has an asymptotic behaviour $\propto t^{-3\ell-m}$~\cite{Price:1971fb,Leaver:1986gd}.
Hence, it is suppressed with respect to the signal observed at $\mathcal{I}^+$ as 
a function of the retarded time $u$, which is instead characterized by a decay $\propto u^{-\ell-2}$.
However, even when observed at finite distance, the late-time decay is characterized by a transient radiative tail $\propto u^{-\ell-2}$, behaving as the one observed at $\mathcal{I}^+$~\cite{Leaver:1986gd}.
This transient eventually leaves place to the $\propto t^{-3\ell-m}$ term.
In Ref.~\cite{Zenginoglu:2008wc}, ID-driven perturbative numerical simulations confirmed this picture, showing that a progressively longer transient appears as the observer is moved further away from the source.
The physical interpretation is the following: tail signals are generated by the interaction of small frequency signals with the long-range, slow decay of the background. 
Smaller frequencies can probe larger scales and get more efficiently back-scattered. 
If the observer is located close to the BH, smaller frequencies cannot reach it and, as consequence, the observed tail is quenched.

\begin{figure}[b!]
\centering
\includegraphics[width=\columnwidth]{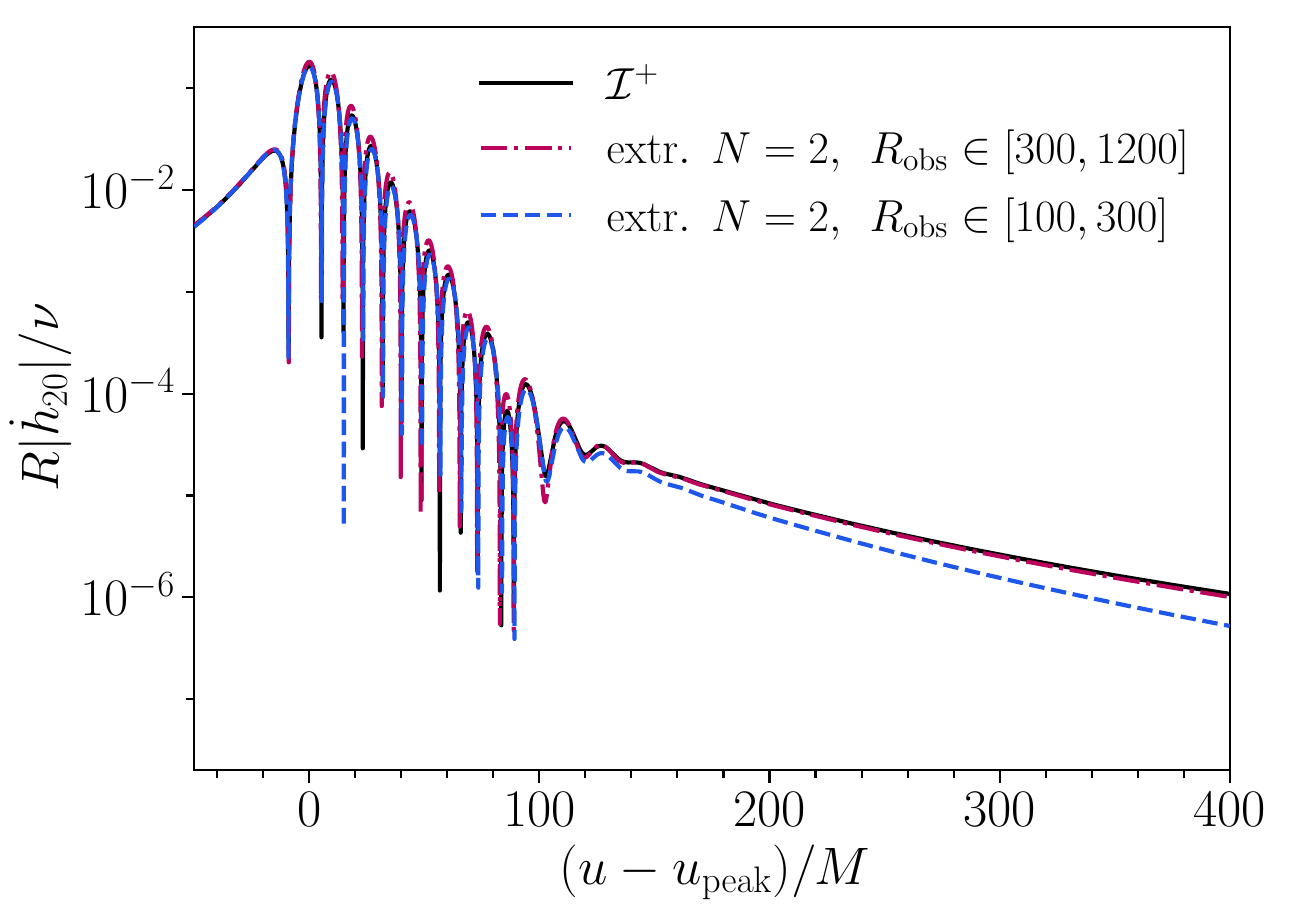}
\caption{
Comparison of the waveform directly extracted at $\mathcal{I}^+$ (black solid) using \textsc{RWZHyp}, while the dashed and dot-dashed lines correspond to waveforms extrapolated to $\mathcal{I}^+$ using a polynomial extrapolation built from finite-distance observer locations $R_{\rm obs}\in[300M,1200M]$ (purple) and $R_{\rm obs}\in[100M,300M]$ (blue).
}
\label{fig:extrap_test_rwz}
\end{figure}

\textit{Extrapolation of nonlinear simulations --} 
With this phenomenology in mind, we extrapolate the numerical
waveforms obtained from nonlinear evolutions to $\mathcal{I}^+$ 
using the SXS standard polynomial extrapolation procedure 
of~\cite{Iozzo21} as implemented in the
\textsc{scri} package~\cite{scri,scrirepo,Boyle2013,BoyleEtAl:2014,Boyle2015a}, with polynomial order $N=2$.
In Fig.~\ref{fig:resolution} we also compare the results obtained with $N=3$, showing that this does not alter our conclusions.
We choose extraction radii at large distances in the interval $[300M, 1200M]$.
To gain intuition about whether our procedure leads to a correct extraction of tails, we perform a test on the perturbative waveforms computed with the \textsc{RWZHyp} code, where we can compare this result directly to waveforms computed at $\mathcal{I}^+$, see Fig.~\ref{fig:extrap_test_rwz}.
In particular, we compare two different extrapolations, computed considering $R_{\rm obs}\in [100M, 300M]$ and $R_{\rm obs}\in [300M,1200M]$ respectively.
As expected from our argument above, the extrapolation performed with $R_{\rm obs}$ closer to the BH yields a large mismatch with respect to the one computed at $\mathcal{I}^+$. 
On the other hand, extrapolating with $R_{\rm obs}$ far enough from the BH, yields an extrapolation in excellent agreement with the tail computed at $\mathcal{I}^+$ by means of the hyperboloidal layer. 
The agreement slowly decreases in time; as expected, the extrapolated waveform undergoes a faster decay after the initial evolution.\\

\noindent {\bf \em Cauchy-characteristic Evolution.} 
%
Apart from extrapolating waveforms to $\mathcal{I}^+$, we also investigate the use of Cauchy-characteristic evolution (CCE).
We run \textsc{SpECTRE} code's CCE module~\cite{spectrecode,Moxon:2020gha,Moxon:2021gbv} on worldtubes with radii
$[300, 600, 900, 1200] M$. 
Initial data for the first null hypersurface for each simulation was created using \textsc{SpECTRE}'s default \textsc{ConformalFactor} method. 
After running CCE, we mapped each system to the superrest frame of its remnant black hole 50M before the end of the simulation~\cite{Mitman:2021xkq,MaganaZertuche:2021syq,Mitman:2022kwt,Mitman:2024uss}.

\begin{figure}[t!]
\centering
\includegraphics[width=\columnwidth]{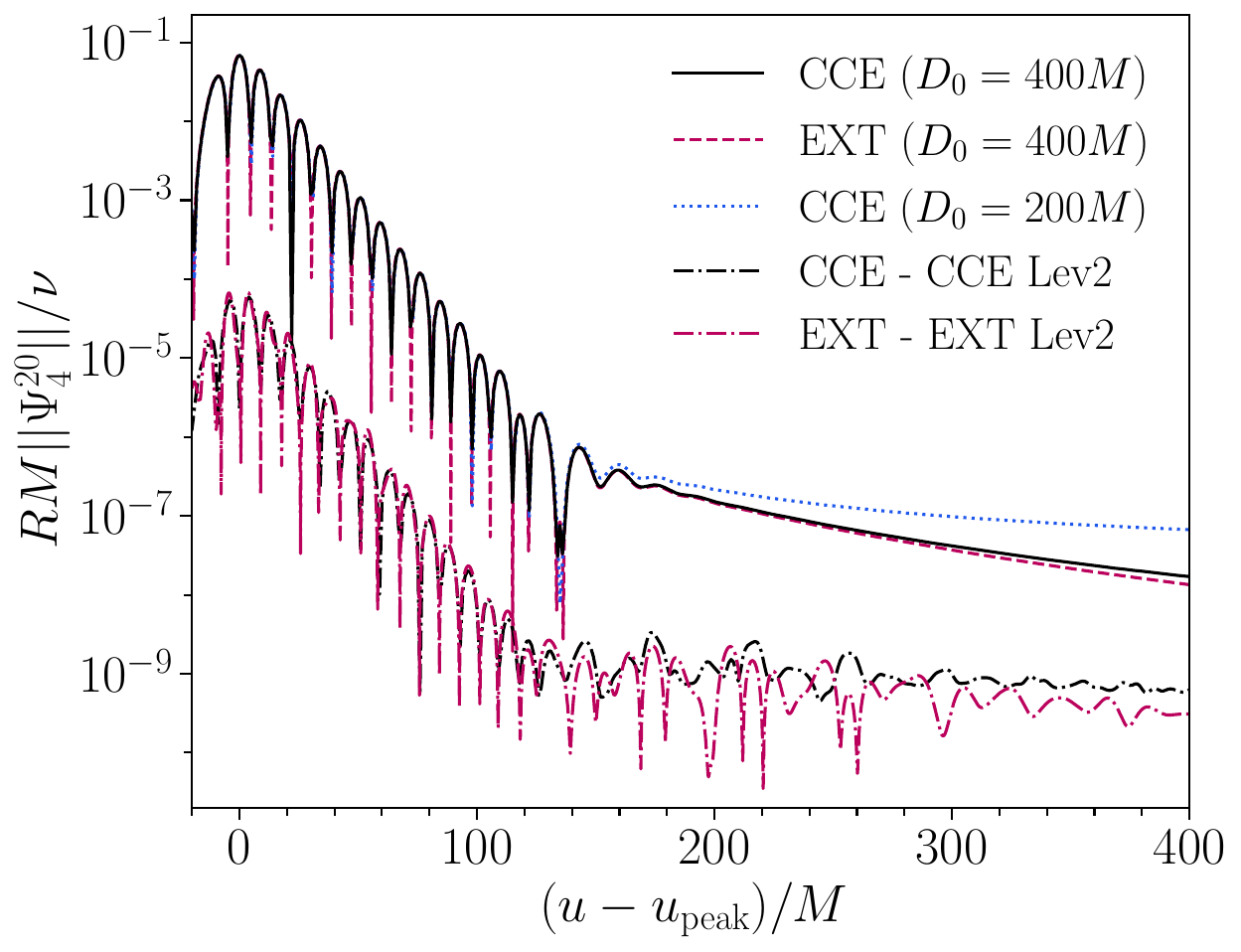}
\caption{Mass-rescaled quadrupolar $\Psi_{4}$ amplitude, as a function of retarded time from the peak.
The CCE ($R_{\text{ext}}=600M$) and extrapolated ($N=2$) waveforms of SXS:BBH:3996 with $D_{0}=400M$ are shown in black and purple.
The CCE waveform SXS:BBH:3995 with $D_{0}=200M$ is shown in blue.
Differences between the two highest Cauchy-evolution resolutions for SXS:BBH:3996 are shown with dot-dashed lines.
}
\label{fig:CCE}
\end{figure}

In Fig.~\ref{fig:CCE} we show the Newman-Penrose scalar $\Psi_{4}$ extracted using CCE (with $R_{\text{ext}}=600M$\footnote{We use $R_{\text{ext}}=600M$ since this worldtube radius shows marginally better agreement with the extrapolated waveform. Larger worldtube radii tend to yield a slightly slower falloff. The reason behind this will be investigated in future work.}) and extrapolation, for runs SXS:BBH:3995 and SXS:BBH:3996.
We also show the residual between the highest and next-highest resolutions for both the CCE and extrapolated waveforms.
As can be seen, while the CCE and extrapolated waveforms for $D_{0}=400M$ seem to agree fairly well, for $D_{0}=200M$ the CCE waveform has a much slower falloff than the extrapolated waveform.
We suspect that this inconsistency is likely due to the initial data in CCE.
In particular, with a shorter simulation, like SXS:BBH:3995 with $D_{0}=200M$, there is less time for the radiation due to unphysical initial data to propagate out of the system.
Consequently, the late-time tail behavior between the two simulations can take on fairly different structures.
We choose to work with extrapolated waveforms in the main text to mitigate this effect.
Understanding the impact of initial data on the tails of CCE waveforms will be the study of future work.\\

\noindent {\bf \em Waveform Filtering.} 
%
\begin{figure}[b!]
\centering
\includegraphics[width=0.9\columnwidth]{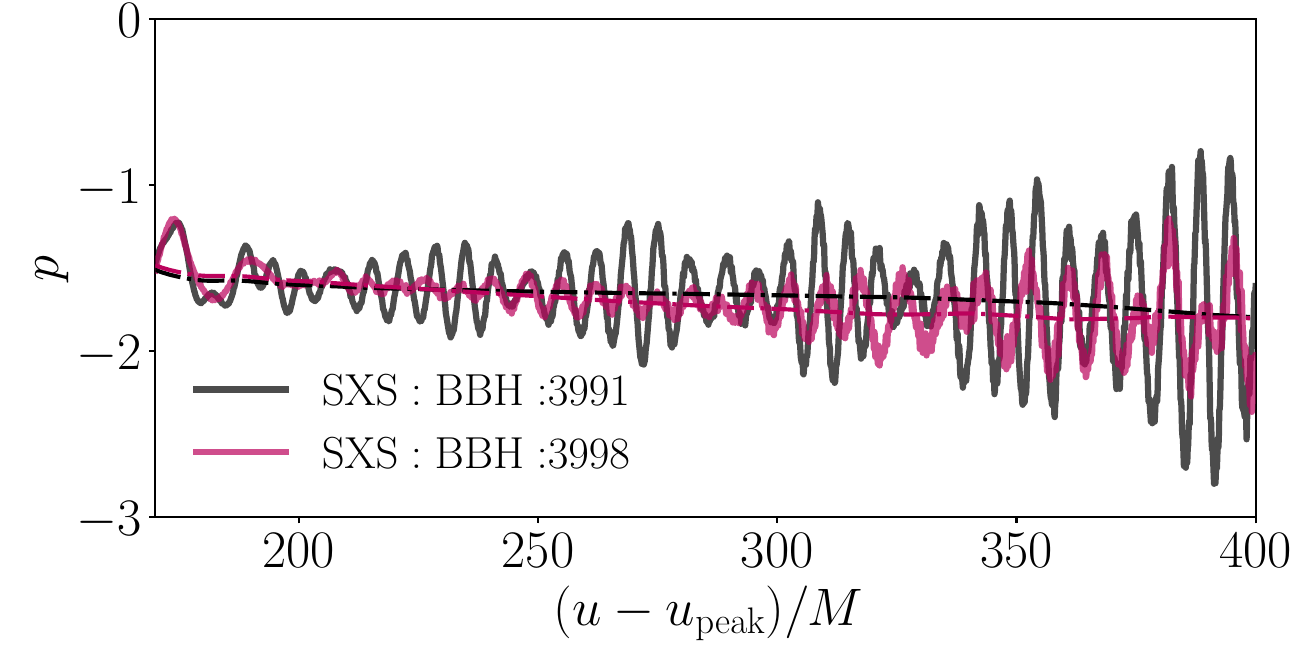}
\caption{
Tail exponent of Eq.~\eqref{eq:tail_exponent} vs time from the peak. 
Results relative to two comparable-mass SXS simulations of Tab.~\ref{tab:sims_ecc}. 
Thick lines are the unfiltered numerical results. Dot-dashed lines are obtained applying a Savitzky-Golay filter both to the news amplitude and to the tail exponent obtained from it.\\
}
\label{fig:p_filter_tests}
\end{figure}
To reduce high-frequency numerical noise in the NR waveforms, we applied a Savitzky-Golay filter. 
The filter, acting on a sliding time window, is applied on the news amplitude in the interval $u-u_{\rm peak}\in[150M,400M]$. 
The two dominant noise frequency components are suppressed by applying the filter with a window length of $20M$ and then $6M$, suppressing the modulations and the high frequency oscillations contaminating the news.
The tail exponent is instead filtered with a $20M$-long window.
The filter fits the data with a polynomial in the specified time window, then set as value of the filtered function the fit prediction at the center of the interval. 
We set the fitting function as linear, using \texttt{scipy.signal.savgol$\_$filter(data, window$\_$length=window$\_$length, polyorder=1)}.

In Fig.~\ref{fig:p_filter_tests} we compare the filtered vs unfiltered tail exponent, for two exemplary \SpEC simulations analyzed in this work.
We report the simulations SXS:BBH:3991 and SXS:BBH:3998, which are evolved under the same initial data, but with different $R_{out}$, to showcase how the result obtained is robust with respect to different binary initial data.
Albeit contaminated by high-frequency noise, a clear decaying trend for the exponent is distinguishable.

\end{document}